\documentclass[conference,a4paper]{IEEEtran}

\usepackage{xcolor}
\usepackage{balance}
\ifCLASSINFOpdf
  \usepackage[pdftex]{graphicx}
\else
  \usepackage[dvips]{graphicx}
\fi
\usepackage{cite}
\usepackage{epstopdf}
\usepackage{multirow}
\usepackage{siunitx}
\usepackage{amsmath}
\usepackage{siunitx}
\usepackage{pgfplots}
\pgfplotsset{compat=newest}
\usepackage{nicefrac}
\newlength\figureheight
\newlength\figurewidth
\newcommand{\figref}{Fig.~\ref}

\usepackage{comment}
\usepackage{dblfloatfix} 
\ifCLASSOPTIONcompsoc
 \usepackage[caption=false,font=normalsize,labelfont=sf,textfont=sf]{subfig}
\else
 \usepackage[caption=false,font=footnotesize]{subfig}
\fi
\usepackage[absolute,overlay]{textpos} 

\begin{document}

\title{Electrically Small Multimodal 3D Beamforming MIMO Antenna for PHY-Layer Security}

\author{\IEEEauthorblockN{
Abel Zandamela\IEEEauthorrefmark{1},   
Nicola Marchetti\IEEEauthorrefmark{1},   
Adam Narbudowicz\IEEEauthorrefmark{1}    
}                                     
\IEEEauthorblockA{\IEEEauthorrefmark{1}%
CONNECT Centre, Trinity College Dublin, The University of Dublin, Dublin, Ireland.\\ \{zandamea, nicola.marchetti, narbudoa\}@tcd.ie}
}

\maketitle

\begin{abstract}
This work proposes an electrically small 3D beamforming antenna for PHYsical Layer (PHY-layer) security. The antenna comprises two layers of stacked patch structures, and is a five-mode five-port MIMO system operating around 1.85 GHz with electrical size $\boldsymbol{ka=0.98}$ and radiation efficiency of up to $\boldsymbol{55\%}$. By studying the properties of the excited modes, phase and amplitude control allow for unidirectional beam scanning towards any direction around the elevation and azimuth planes. PHY-layer security is investigated using the directional modulation (DM) technique, which transmits unscrambled baseband constellations symbols to a pre-specified secure direction, while simultaneously spatially distorting the same constellations in all other directions. Bit Error Rate (BER) calculations reveal very low values of $\boldsymbol{2\times10^{-5}}$ for the desired direction of the legitimate receiver, with BER$\boldsymbol{<10^{-2}}$ beamwidths of $\boldsymbol{55^{\circ}}$ and $\boldsymbol{58^{\circ}}$ for the azimuth and elevation planes, respectively.

\end{abstract}

\vskip0.5\baselineskip

\section{Introduction}

\begin{figure}[!ht]
\centering
{\includegraphics[width=1.0\columnwidth]{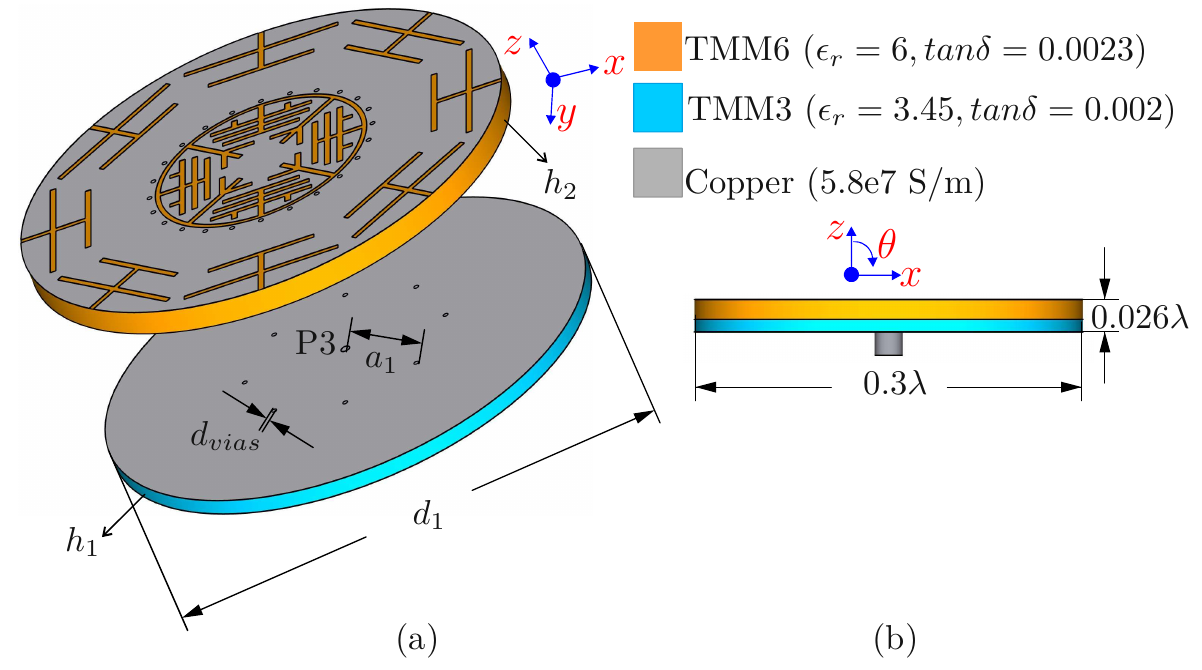}}
\hfil
{\includegraphics[clip, trim=0cm 0cm 0.6cm 0cm, width=\columnwidth]{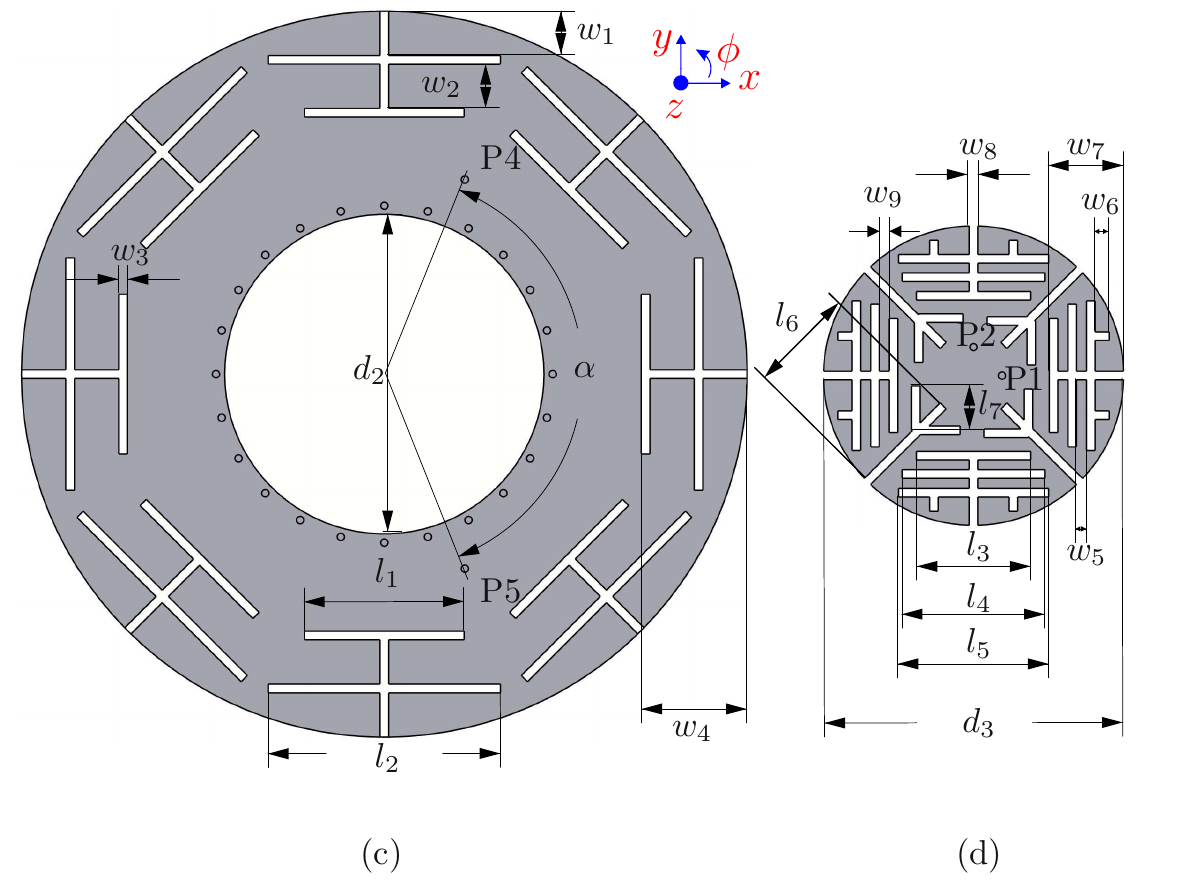}}
\caption{Proposed 3D beamforming electrically small antenna: (a) exploded view showing the two stacked layers and the configuration of the bottom layer; (b) side view highlighting antenna diameter and thickness; (c) top-view of the shorted-ring patch with port arrangement and (d) top-view of the central patch; antenna dimensions (all in mm):~$a_1=10.4$, $d_1=50$, $d_2=22$, $d_3=21$, $d_{vias}=0.5$, $h_1=1.57$, $h_2=2.54$, $l_1=11$, $l_2=16$, $l_3=8$, $l_4=10$, $l_5=10.5$, $l_6=7.5$, $w_1=3.05$, $w_2=3.05$, $w_3=0.6$, $w_4=7.3$, $w_5=0.7$, $w_6=1$, $w_7=5.2$, $w_8=0.6$, $w_9=0.7$, $\alpha=\ang{135}$; feed locations from the disk center (in mm): P$1=$~P$2=2$, P$3=0$, P$4=$~P$5=3.5$.}
\label{fig:system_setup}
\end{figure}

Antenna radiation pattern reconfigurability is a highly sought-after feature to enable many cutting-edge systems like smart cities and the Internet of Things (IoT) \cite{Qadri2020, Dian2020}. Pattern control is essential for localization using triangulation via relative angles measurements \cite{Munoz2009, Obeidat2021, IoTMAG22} and emerging wireless cryptography methods based on physical radio propagation \cite{Trung2018, Wei2020, Daly2009, Ding2014, AWPL2022}. However, modern technological trends pose size constraints in many IoT applications, leading to the high demand for miniaturized antennas.

In recent years different compact pattern reconfigurable antennas have been proposed in the open literature, e.g. \cite{Iqbal2021, Zhao2022, Huang2021, Wu2020}. A single-layer dual-mode circular patch antenna is investigated for self-scanning and nulling properties in \cite{Iqbal2021}; the design has a total size of $ka = 4.2$ (where $a$ is
the radius of the smallest sphere that can completely enclose the
antenna at the centre operating frequency $f_0=1/\lambda_0$, and $k =
2\pi/\lambda_0 $ is the free space wavenumber). Modified array structures ($ka=0.98$) controlled by PIN diodes are used to switch between unidirectional and omnidirectional patterns in \cite{Zhao2022}. In \cite{Huang2021}, a single-layer pattern reconfigurable antenna ($ka=0.75$) comprising six PIN diodes for three switchable endfire states, is investigated for full-azimuth plane coverage. An ultra-thin pattern reconfigurable
metamaterial-inspired Huygens dipole antenna ($ka=0.98$), using two PIN diodes to switch between two unidirectional endfire states, is proposed in \cite{Wu2020}. However, in the above works, some structures either still have a relatively large electrical size \cite{Iqbal2021}, or due to the integration of externally controlled PIN diodes, the beamsteering performance of such designs are restricted to a few discreet states that cannot be activated simultaneously\cite{Zhao2022, Huang2021, Wu2020}; which limits their use for advanced applications like localization, PHY-layer security, and frequency-division multiple access systems. 

In this work, we investigate for the first time the feasibility of electrically small 3D beamforming antennas for PHY-layer security via directional modulation (DM). The proposed antenna exhibits total efficiency of up to $55\%$ and is capable of unidirectional beamforming in the azimuth and elevation planes. DM performance shows that the antenna realizes secure transmissions with BER$<10^{-2}$ of $\ang{55}$ and $\ang{58}$ beamwidths for either azimuth or elevation plane.

\section{Antenna Design and Working Principle}

\subsection{Antenna Configuration}

The proposed electrically small antenna is shown in \figref{fig:system_setup}. It comprises two stacked layers of diameter $d_1=\SI{50}{mm}$, with thickness $h_1=\SI{1.57}{mm}$ (bottom-layer) and $h_2=\SI{2.54}{mm}$ (top-layer). The top layer consists of a central circular patch of diameter $d_3 = \SI{21}{mm}$ (see \figref{fig:system_setup}d) and a concentric shorted-ring patch (shown in \figref{fig:system_setup}c), and is supported by a TMM6 substrate ($\epsilon_r=6$ and $tan \delta = 0.0023$). The shorted-ring patch excites two orthogonal TM$_{21}$ modes and is fed using ports P4 and P5, rotated by $\alpha=\ang{135}$ and located at $\SI{3.5}{mm}$ from its inner edges. A total of 24 shorting-pins ($d_{vias}=\SI{0.5}{mm}$) rotated by $\ang{15}$ and located $\SI{0.6}{mm}$ from the inner edge, are introduced in the ring for frequency control and isolation enhancement. For further miniaturization, eight double T-shaped slits rotated by $\ang{45}$ are present in the concentric ring; the double T-shaped slits have different lengths ($l_1$ for the top T-shaped and $l_2$ for the bottom T-shaped slits), and the width is $w_3=\SI{0.6}{mm}$. 

The central patch (\figref{fig:system_setup}d) excites two orthogonal TM$_{11}$ broadside modes and is fed using ports P1 and P2 rotated by $\ang{90}$, and located at $\SI{2}{mm}$ from the disk center. To increase the current path, four triple T-shaped slits are rotated by $\ang{90}$ in the patch. The width of the slits is $w_8=\SI{0.6}{mm}$ and the slits have different lengths as shown in (\figref{fig:system_setup}d); in addition, diagonal slits of length $l_6$ are also used for further miniaturization. 

The bottom layer comprises a shorted-pin circular patch fed using port P3 at the patch center, to excite a monopole-like pattern. The patch uses TMM3 substrate ($\epsilon_r=3.45$ and $tan \delta=0.002$) and includes a total of eight pins, rotated by $\ang{45}$ and located at $a_1=\SI{10.4}{mm}$ from the disk center. 

\subsection{Working Principle}

\begin{figure}[!ht]
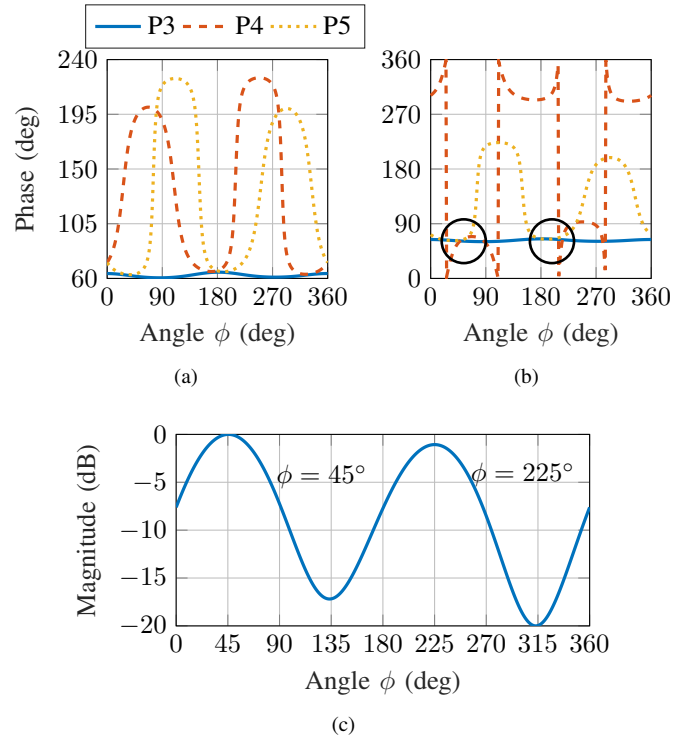

\centering
\subfloat[]{
\hspace{-0.5cm}
    \setlength\figureheight{0.16\textwidth}
	\setlength\figurewidth{0.16\textwidth}
	\input{Phase_NoShift}}
\hfill
\subfloat[]{
    \hspace{-0.5cm}
    \setlength\figureheight{0.16\textwidth}
	\setlength\figurewidth{0.16\textwidth}
	\input{Phase_45deg}}
	\hfill
\subfloat[]{
    \hspace{-0.5cm}
    \setlength\figureheight{0.14\textwidth}
	\setlength\figurewidth{0.3\textwidth}
%
%
\definecolor{mycolor1}{rgb}{0.00000,0.44700,0.74100}%
\begin{tikzpicture}

\begin{axis}[%
width=\figurewidth,
height=\figureheight,
at={(0,0)},
scale only axis,
xmin=0,
xmax=360,
xtick={  0,  45,  90, 135, 180, 225, 270, 315, 360},
xlabel style={font=\color{white!15!black}},
xlabel={$\text{Angle }\phi\text{ (deg)}$},
ymin=-20,
ymax=0,
ylabel style={font=\color{white!15!black}},
ylabel={Magnitude (dB)},
axis background/.style={fill=white},
xmajorgrids,
ymajorgrids
]
\addplot [color=mycolor1, line width=1.2pt, forget plot]
  table[row sep=crcr]{%
0	-7.61927452318404\\
1	-7.30966460074078\\
2	-7.00468085174486\\
3	-6.70451350267527\\
4	-6.40935568222356\\
5	-6.11939030808122\\
6	-5.83479053295391\\
7	-5.55572017233148\\
8	-5.28233411486925\\
9	-5.01477871621938\\
10	-4.75319217712612\\
11	-4.49770490655964\\
12	-4.24843987062096\\
13	-4.00551292790546\\
14	-3.76903315196348\\
15	-3.53910314144601\\
16	-3.31581931847494\\
17	-3.09927221573176\\
18	-2.88954675270425\\
19	-2.68672250149227\\
20	-2.49087394252527\\
21	-2.30207071050259\\
22	-2.12037783082615\\
23	-1.94585594676345\\
24	-1.77856153752595\\
25	-1.61854712743726\\
26	-1.4658614863049\\
27	-1.32054982110585\\
28	-1.18265395904196\\
29	-1.05221252201365\\
30	-0.929261092519528\\
31	-0.813832370975494\\
32	-0.705956324418693\\
33	-0.605660326538274\\
34	-0.512969288966815\\
35	-0.42790578373052\\
36	-0.350490156755541\\
37	-0.280740632303168\\
38	-0.218673408194913\\
39	-0.164302741676334\\
40	-0.117641025755876\\
41	-0.078698855843637\\
42	-0.0474850865066045\\
43	-0.0240068781457516\\
44	-0.00826973339281878\\
45	-3.61424185335579e-05\\
46	0\\
47	-0.00748131931037221\\
48	-0.0227087929422041\\
49	-0.04567949229288\\
50	-0.0763888177981791\\
51	-0.114830455827175\\
52	-0.160996323373034\\
53	-0.214876500243822\\
54	-0.276459148438061\\
55	-0.345730418372899\\
56	-0.422674341609397\\
57	-0.50727270969761\\
58	-0.59950493873836\\
59	-0.699347919231297\\
60	-0.80677585075199\\
61	-0.921760060969124\\
62	-1.04426880848493\\
63	-1.17426706895038\\
64	-1.31171630387795\\
65	-1.45657421154771\\
66	-1.6087944593712\\
67	-1.76832639706939\\
68	-1.93511474998646\\
69	-2.10909929186889\\
70	-2.29021449642774\\
71	-2.478389167017\\
72	-2.67354604378361\\
73	-2.87560138768772\\
74	-3.08446454084707\\
75	-3.3000374627468\\
76	-3.52221424196109\\
77	-3.75088058317964\\
78	-3.98591326950418\\
79	-4.22717960019746\\
80	-4.47453680433492\\
81	-4.7278314311204\\
82	-4.98689871800227\\
83	-5.25156193816584\\
84	-5.52163172947844\\
85	-5.79690540755393\\
86	-6.07716626626461\\
87	-6.36218286977324\\
88	-6.65170834100731\\
89	-6.94547965242116\\
90	-7.24321300310601\\
91	-7.54461249713195\\
92	-7.84936483906317\\
93	-8.15713553435848\\
94	-8.46757070922376\\
95	-8.78029660073115\\
96	-9.0949191232678\\
97	-9.41102352720951\\
98	-9.72817416698608\\
99	-10.0459143968851\\
100	-10.3637666139236\\
101	-10.6812324678492\\
102	-10.9977932587495\\
103	-11.3129105427404\\
104	-11.6260269656992\\
105	-11.9365673439189\\
106	-12.2439400087799\\
107	-12.5475384300121\\
108	-12.8467431287743\\
109	-13.1409238875606\\
110	-13.4294422588741\\
111	-13.7116543686644\\
112	-13.98691400383\\
113	-14.254575965699\\
114	-14.5139996635623\\
115	-14.7645529141958\\
116	-15.0056159052209\\
117	-15.2365852723808\\
118	-15.4568782337517\\
119	-15.6659367179026\\
120	-15.8632314184456\\
121	-16.0482657046172\\
122	-16.2205793167722\\
123	-16.3797517771563\\
124	-16.5254054501772\\
125	-16.657208192569\\
126	-16.7748755422313\\
127	-16.8781724048442\\
128	-16.9669142092493\\
129	-17.0409675155562\\
130	-17.1002500734599\\
131	-17.1447303417586\\
132	-17.1744264929702\\
133	-17.1894049387168\\
134	-17.1897784217594\\
135	-17.1753679815891\\
136	-17.1473706546281\\
137	-17.1049935109356\\
138	-17.048874822131\\
139	-16.9793187585488\\
140	-16.8966586029552\\
141	-16.801253505091\\
142	-16.6934852813052\\
143	-16.57375530411\\
144	-16.4424815191752\\
145	-16.3000956197058\\
146	-16.1470404006674\\
147	-15.9837673081751\\
148	-15.8107341927882\\
149	-15.6284032695815\\
150	-15.437239282833\\
151	-15.2377078689913\\
152	-15.0302741082942\\
153	-14.815401252962\\
154	-14.5935496182305\\
155	-14.3651756215158\\
156	-14.1307309546666\\
157	-13.8906618743988\\
158	-13.6454085965811\\
159	-13.3954047808982\\
160	-13.1410770935159\\
161	-12.882844836584\\
162	-12.6211196347186\\
163	-12.3563051698921\\
164	-12.0887969574334\\
165	-11.8189821570296\\
166	-11.5472394137071\\
167	-11.2739387247621\\
168	-10.9994413294537\\
169	-10.7240996190083\\
170	-10.4482570650933\\
171	-10.1722481654061\\
172	-9.8963984054094\\
173	-9.62102423553162\\
174	-9.346433063362\\
175	-9.07292326049332\\
176	-8.80078418375474\\
177	-8.53029621059229\\
178	-8.26173078835838\\
179	-7.99535049722966\\
180	-7.73146797752665\\
181	-7.47020455262462\\
182	-7.21186193615569\\
183	-6.9566680103465\\
184	-6.70484229223003\\
185	-6.45659606812232\\
186	-6.21213253799586\\
187	-5.97164696901863\\
188	-5.73532685747948\\
189	-5.50335209831403\\
190	-5.27589516141954\\
191	-5.05312127395089\\
192	-4.83518860779234\\
193	-4.62224847140427\\
194	-4.41444550527571\\
195	-4.21191788022076\\
196	-4.01479749779098\\
197	-3.82321019210913\\
198	-3.63727593245187\\
199	-3.45710902595257\\
200	-3.28281831981938\\
201	-3.11450740251075\\
202	-2.95227480332637\\
203	-2.79621418992263\\
204	-2.64641456327967\\
205	-2.50296044967832\\
206	-2.36593208928001\\
207	-2.23540562091669\\
208	-2.11145326272898\\
209	-1.99414348830713\\
210	-1.8835411980117\\
211	-1.77970788515793\\
212	-1.68270179677269\\
213	-1.59257808863961\\
214	-1.50938897434947\\
215	-1.43318386810109\\
216	-1.36400952097793\\
217	-1.30191015045251\\
218	-1.24692756286071\\
219	-1.19910126859335\\
220	-1.15846858975328\\
221	-1.1250647600237\\
222	-1.09892301649452\\
223	-1.08007468318429\\
224	-1.06854924600123\\
225	-1.06412176925744\\
226	-1.06755778464171\\
227	-1.0781404972623\\
228	-1.09614629940908\\
229	-1.12159599602207\\
230	-1.15450880668693\\
231	-1.19490237160778\\
232	-1.24279274666495\\
233	-1.2981943872053\\
234	-1.36112012019176\\
235	-1.43158110431396\\
236	-1.5095867776332\\
237	-1.59514479231047\\
238	-1.68826093593227\\
239	-1.78893903890554\\
240	-1.89718086736572\\
241	-2.01298600098404\\
242	-2.13635169502252\\
243	-2.26727272592609\\
244	-2.40574121968898\\
245	-2.55174646216605\\
246	-2.70527469044089\\
247	-2.86630886428779\\
248	-3.03482841669381\\
249	-3.21080898233294\\
250	-3.39422210279961\\
251	-3.58503490733465\\
252	-3.78320976768739\\
253	-3.98870392568537\\
254	-4.20146909199291\\
255	-4.421451014474\\
256	-4.64858901450076\\
257	-4.88281548948842\\
258	-5.12405537989964\\
259	-5.37222559891685\\
260	-5.62723442299469\\
261	-5.88898084150446\\
262	-6.15735386375851\\
263	-6.43223178178777\\
264	-6.71348138740739\\
265	-7.00095714231229\\
266	-7.29450030024304\\
267	-7.59393798063267\\
268	-7.89908219363209\\
269	-8.20972881701362\\
270	-8.52565652619104\\
271	-8.84663525668642\\
272	-9.17238096071394\\
273	-9.50262922413471\\
274	-9.83707831590642\\
275	-10.1754033696517\\
276	-10.5172551225422\\
277	-10.8622586528673\\
278	-11.2100121309489\\
279	-11.5600856009472\\
280	-11.9120198142686\\
281	-12.265325138725\\
282	-12.6194805712603\\
283	-12.9739328858858\\
284	-13.3280959523914\\
285	-13.6813502652701\\
286	-14.0330427259883\\
287	-14.3824867250477\\
288	-14.7289625729967\\
289	-15.0717183313729\\
290	-15.4099710952178\\
291	-15.7429087779441\\
292	-16.0696924466113\\
293	-16.3894592507565\\
294	-16.7013259804755\\
295	-17.0043932792488\\
296	-17.2977505238445\\
297	-17.5804813675284\\
298	-17.851669923848\\
299	-18.1104075468614\\
300	-18.3558001403596\\
301	-18.5869759042618\\
302	-18.8030934019979\\
303	-19.0033498095866\\
304	-19.1869891867017\\
305	-19.3533105937489\\
306	-19.5016758682479\\
307	-19.6315168698865\\
308	-19.7423420073498\\
309	-19.833741871976\\
310	-19.9053938233527\\
311	-19.9570653996173\\
312	-19.9886164592558\\
313	-20\\
314	-19.9912616419728\\
315	-19.9621342038033\\
316	-19.9139979862372\\
317	-19.8460182277555\\
318	-19.7589699392863\\
319	-19.6533110305071\\
320	-19.5295650735207\\
321	-19.3883141477285\\
322	-19.2301913911548\\
323	-19.0558734336751\\
324	-18.8660728742241\\
325	-18.6615309460108\\
326	-18.4430104925245\\
327	-18.2112893540666\\
328	-17.9671542410385\\
329	-17.711395147353\\
330	-17.4448003360494\\
331	-17.1681519101225\\
332	-16.8822219651604\\
333	-16.5877693068142\\
334	-16.2855367054495\\
335	-15.9762486523875\\
336	-15.6606095767433\\
337	-15.3393024786914\\
338	-15.0129879336928\\
339	-14.682303422487\\
340	-14.3478629431026\\
341	-14.0102568635251\\
342	-13.6700519766351\\
343	-13.3277917224221\\
344	-12.9839965460468\\
345	-12.6391643639288\\
346	-12.2937711135723\\
347	-11.9482713661819\\
348	-11.6030989842532\\
349	-11.2586678091673\\
350	-10.9153723663845\\
351	-10.5735885781059\\
352	-10.2336744752551\\
353	-9.89597090235106\\
354	-9.56080221029612\\
355	-9.22847693335346\\
356	-8.89928844759178\\
357	-8.57351560893867\\
358	-8.25142336965963\\
359	-7.93326337262802\\
360	-7.61927452318404\\
};
\end{axis}

\begin{axis}[%
width=\figurewidth,
height=\figureheight,
at={(0in,0in)},
scale only axis,
xmin=0,
xmax=1,
ymin=0,
ymax=1,
axis line style={draw=none},
ticks=none,
axis x line*=bottom,
axis y line*=left
]
\node[below right, align=left]
at (rel axis cs:0.22,0.89) {$\phi =45^{\circ}$};
\node[below right, align=left]
at (rel axis cs:0.69,0.9) {$\phi =225^{\circ}$};
\end{axis}
\end{tikzpicture}%
}
\caption{Phase properties of the modes used for beamsteering in the azimuth plane: (a) phase patterns without phase shift; (b) phase patterns with phase shifts in P4 and P5 for beamsteering towards $\phi=\ang{45}$; and (c) normalized synthesized patterns for $\phi=\ang{45}$ direction. Note that this performance has an ambiguity because a second main beam is also present at $\phi+\ang{180}$ direction.}
\label{fig:Phase_properties}
\end{figure}

The beamsteering principle is based on phase and amplitude control of the different feeding ports of the proposed antenna. By studying the phase patterns of each excited mode, phase shifts can be introduced to create constructive interference towards the desired directions. To demonstrate this principle, beamsteering is first analyzed for the azimuth plane. \figref{fig:Phase_properties}a shows the phase patterns (without any phase shift) of each port required for azimuth plane scanning (P3, P4, and P5). Since P3 has a constant phase around this plane, it is used as the reference port and phase shift ($\Delta_{phase(P_{n})}=Phase_{P3} - Phase_{P_{n}}$) is introduced in P4 and P5 to steer the beam towards the desired angle $\phi$.

\figref{fig:Phase_properties}b shows the phase patterns for an exemplar angle $\phi=\ang{45}$. It is seen that by introducing $\Delta_{phase(P_{n})}$, all the three ports are in-phase, and the generated beam pattern is shown in \figref{fig:Phase_properties}c. Because a second in-phase state is also observed around $\phi=\ang{225}$, the obtained pattern will have a bi-directional characteristic. This beamsteering performance will also be observed for different angles, where a second main beam will always be seen at $\phi + \ang{180}$. This performance can compromise the secrecy of the transmitted signals as unwanted users may exploit the ambiguity to retrieve the transmitted information. As it will be shown in the following Sections, the ambiguity can be eliminated by exciting different orthogonal modes. It should also be noted that a closed-form expression can be derived for the beamsteering around the azimuth plane (see \cite{AbelAEU} for more details). 

\section{Full-Wave Simulations and  Discussion}

\subsection{Beamsteering Performance}

\begin{figure}[!ht]
	\setlength\figureheight{0.16\textwidth}
	\setlength\figurewidth{0.37\textwidth}
	\input{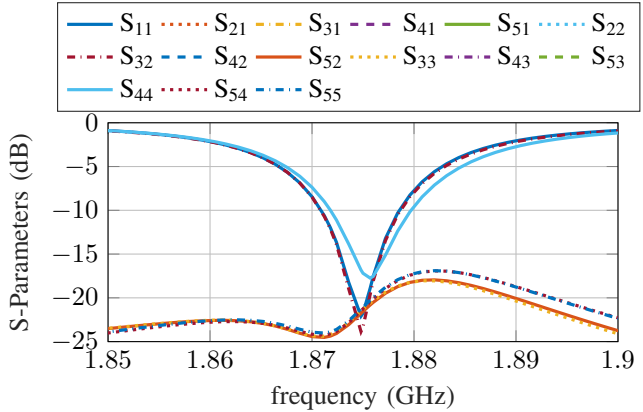}
	\caption{Simulated S-parameters of the proposed  antenna.}
    \label{fig:Spara}
\end{figure}

The proposed electrically small 3D beamforming MIMO antenna is simulated using the finite element method in CST Studio Suite. The antenna S-parameters are shown in \figref{fig:Spara}, the center operating frequency is $f_0=\SI{1.875}{GHz}$, where the isolation is better than $\SI{20.4}{dB}$ between all the antenna ports. The overlapping $\SI{-10}{dB}$ bandwidth between all the antenna ports is $\SI{4.8}{MHz}$. The antenna total efficiency at $f_0$ is around $15\%$ for ports P4 and P5, $30\%$ for ports P1 and P2 and $55\%$ for port P3. Note that the relatively low efficiency of the ports P4 and P5 is due to increased phase variations of the excited modes (TM$_{21}$), which require a much larger diameter to support the dual-phase variations around the antenna perimeter, as discussed in \cite{Abel2019}.   

\subsubsection{Elevation and Azimuth Planes Beamforming}

\begin{figure}[!b]
\centering
{\includegraphics[clip, trim=3cm 8.6cm 2cm 8cm,width=.9\columnwidth]{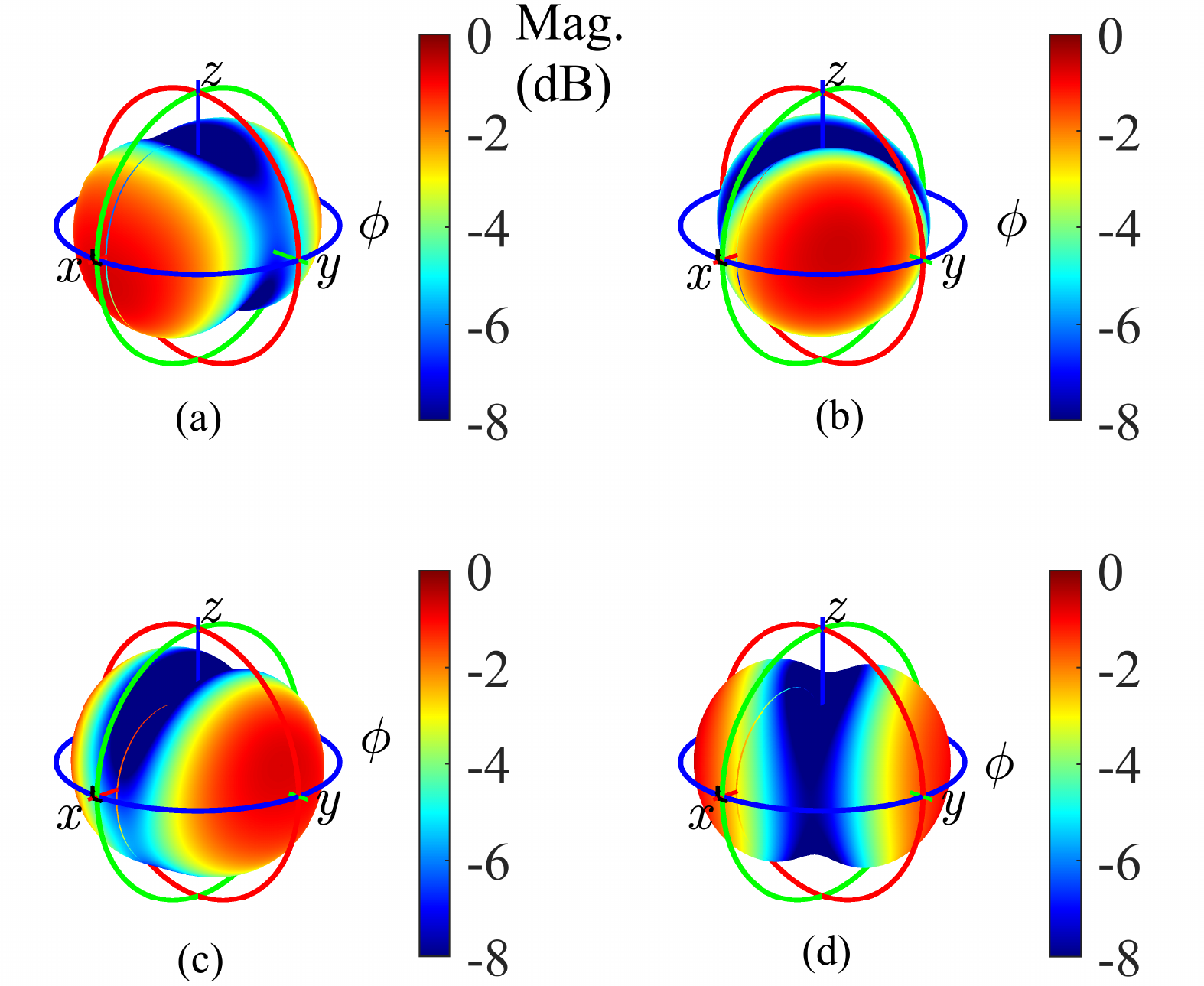}}
\caption{Bidirectional azimuth beamsteering  (obtained by phase control of ports P3, P4, and P5) for four different directions covering the entire azimuth plane: (a) $\ang{0}/\ang{180}$; (b)  $\ang{45}/\ang{225}$; (c)  $\ang{90}/\ang{270}$; and (d) $\ang{135}/\ang{315}$.}
\label{fig:AziBeamsteerPerf}
\end{figure}

The principle described in Section II.B is used to realize the beamforming in the azimuth plane. The performance is shown in \figref{fig:AziBeamsteerPerf} for four different angles to cover the entire plane.

\begin{figure}[!b]
\centering
{\includegraphics[clip, trim=3cm 8.6cm 2cm 8cm,width=.9\columnwidth]{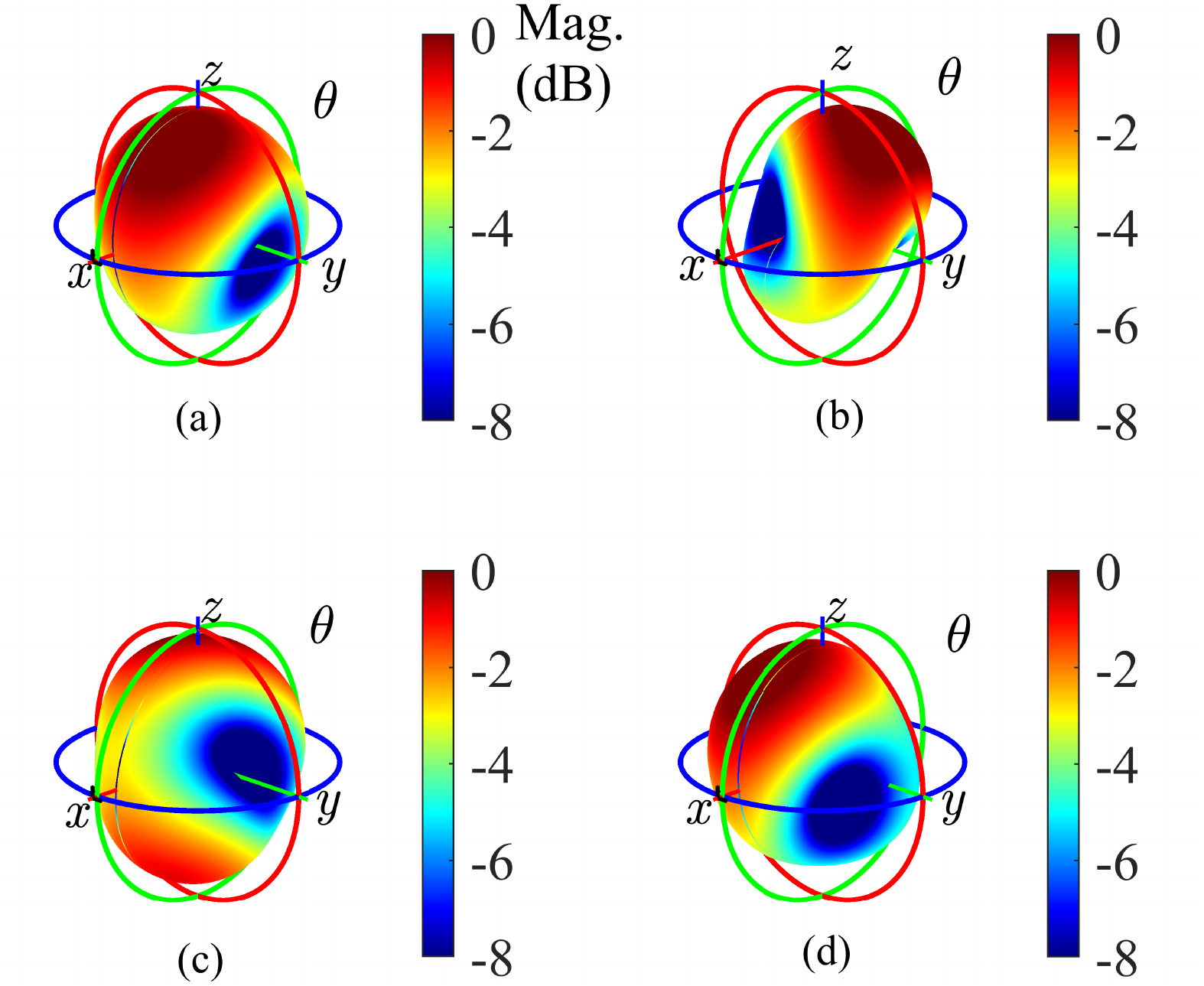}}
\caption{Proposed antenna beamsteering performance for the elevation planes (obtained by combining the modes excited in the top-layer using ports P1, P2, P4, and P5) for four different directions: (a) $\phi=\ang{0},\theta=\ang{22}$; (b) $\phi=\ang{90},\theta=\ang{28}$; (c) $\phi=\ang{0},\theta=\ang{-22}$; and (d) $\phi=\ang{90},\theta=\ang{-28}$.}
\label{fig:ElevBeamsteerPerf}
\end{figure}

The beamsteering in the elevation plane is obtained by combining only the modes excited by the top-layer [TM$_{21}$ modes excited by the shorted-ring patch (P4 \& P5), and the broadside TM$_{11}$ modes excited by the circular patch (P1 \& P2)]. \figref{fig:ElevBeamsteerPerf} shows the beamsteering performance for four different directions: $xz$-plane $\theta=\ang{22}$ obtained by combining P1 and P4 (\figref{fig:ElevBeamsteerPerf}a),  $xz$-plane $\theta=-\ang{22}$ derived from P1 and P5 (\figref{fig:ElevBeamsteerPerf}c), $yz$-plane $\theta=\ang{28}$ (\figref{fig:ElevBeamsteerPerf}b) obtained from P2 and P5, and $yz$-plane $\theta=-\ang{28}$ from P2 and P4 (\figref{fig:ElevBeamsteerPerf}d).

\begin{figure*}[!ht]
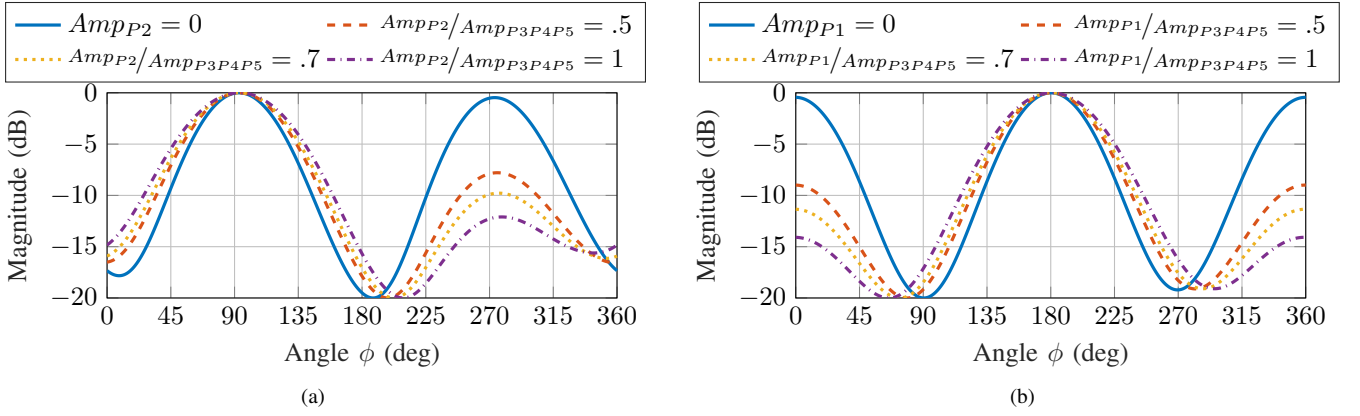

    \centering
	\subfloat[]{
	\setlength\figureheight{0.15\textwidth}
	\setlength\figurewidth{0.37\textwidth}
	\input{90degBeamCompar}
	\label{fig:90degEnhancement}}
	\hfill
	\subfloat[]{
	\hspace{-0.6cm}
	\setlength\figureheight{0.15\textwidth}
	\setlength\figurewidth{0.37\textwidth}
	\input{180degBeamCompar}
	\label{fig:180degEnhancement}}
	\caption{Antenna directivity analysis around the azimuth-plane using different amplitude ratios between P1/P2 and P3P4P5: (a) $\phi=\ang{90}$; and (b) $\phi=\ang{180}$.}
    \label{fig:Directivity_enhancement}
\end{figure*}

\begin{figure}[!ht]
\centering
{\includegraphics[clip, trim=3cm 8.6cm 2cm 8cm,width=.9\columnwidth]{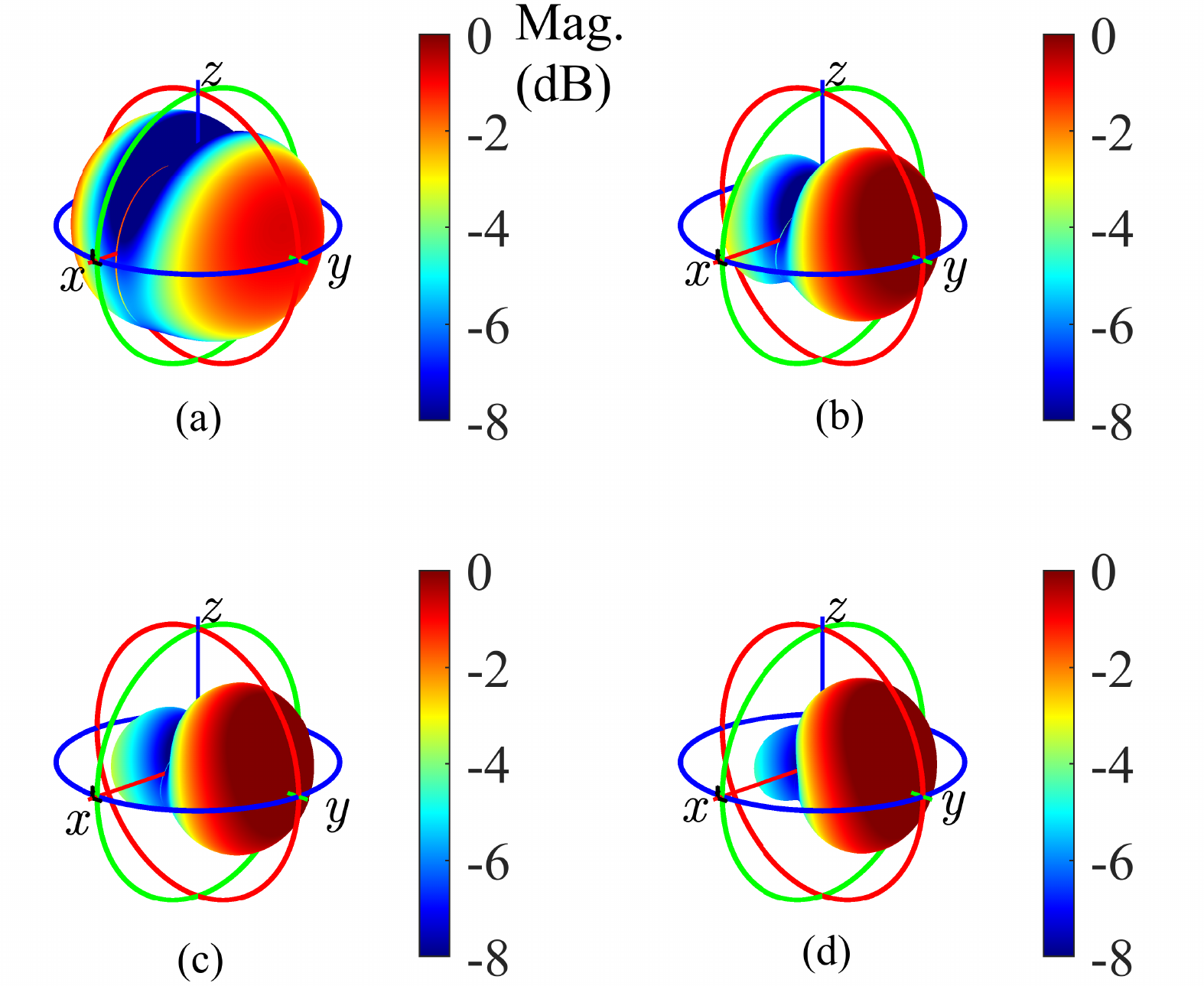}}
\caption{3D visualization of the directivity enhancement for $\phi=\ang{90}$ direction shown in \figref{fig:90degEnhancement}: (a) $Amp_{P2}=0$; (b)  $Amp_{P2}/Amp_{P3P4P5}=.5$; (c)  $Amp_{P2}/Amp_{P3P4P5}=.7$; and (d) $Amp_{P2}/Amp_{P3P4P5}=1$.}
\label{fig:3D90degDirEnhancement}
\end{figure}

\subsubsection{Azimuth-plane Directivity Enhancement}

To eliminate the ambiguity in the beamsteering performance around the azimuth plane, we investigate the simultaneous excitation of the omnidirectional modes (P3, P4 and P5) and a broadside mode (P1/P2). To enhance the directivity, the excited broadside mode is chosen according to the desired $\phi$ direction. In other words, the mode where the farfield pattern has a strong contribution towards the desired $\phi$ angle is selected. Next, the phase pattern of the broadside mode is analyzed to compute the phase shift required to create the in-phase state (similar to the method used in Section II.B). Lastly, the amplitude ratio between the port of the selected broadside mode (P1/P2) and the modes used for the bi-directional pattern in the azimuth plane (P3 P4 P5), is used to control the broadside modes' contribution to the generated farfield patterns. This is required to ensure that the maximum directivity is around the azimuth plane.
\figref{fig:90degEnhancement} illustrates the above-described method to enhance the directivity in the $\phi=\ang{90}$ direction; the respective 3D patterns are shown in \figref{fig:3D90degDirEnhancement}. In this case, P2 is used to reduce the magnitude of the second main beam located at $\phi+\ang{180}$. Different amplitude ratios are used to evaluate the proposed method, revealing that higher amplitude ratios have higher reductions of the $\phi+\ang{180}$ beam magnitude; however, higher amplitude ratios might also increase the contribution toward the elevation directions, for this reason, an amplitude ratio between 0.5 and 0.7 is recommended. 

\figref{fig:180degEnhancement} shows the performance of the proposed method for a different direction, $\phi=\ang{180}$. For this angle, the fields of P1 will strongly contribute towards $\phi=\ang{180}$; therefore, the amplitude ratios of P1 and (P3 P4 P5) are studied to provide the unidirectional performance. Similar to \figref{fig:90degEnhancement}, the $\phi+\ang{180}$ magnitude changes for different amplitude ratios.    

\begin{figure}[!ht]
    \centering
	\setlength\figureheight{0.2\textwidth}
	\setlength\figurewidth{0.37\textwidth}
	\input{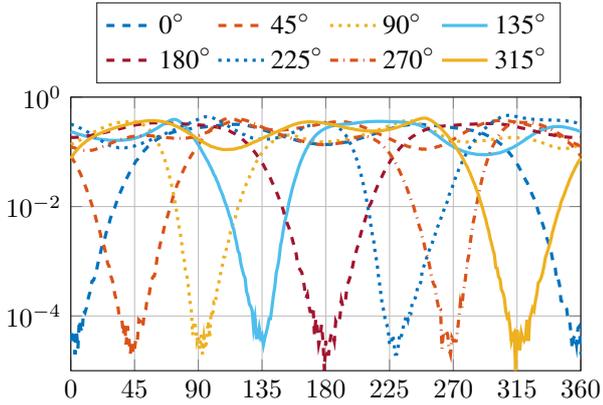}
	\caption{BER calculations (using $\SI{12}{dB}$ SNR) of the proposed antenna for eight different directions separated by $\ang{45}$, covering the entire azimuth-plane.}
    \label{fig:DM_over_xy}
\end{figure}

\section{Directional Modulation}

This Section investigates the beamsteering performance of the  proposed electrically small antenna to realize directional modulation (DM). DM is a physical-layer security method proposed to transmit baseband constellation symbols to a pre-specified secure direction of the legitimate receiver, while simultaneously spatially distorting the same symbols in all the other directions. For this study the DM is investigated for both the azimuth and elevation planes, and the directionally modulated signals are given by

\begin{subequations}
\begin{align}
 s_{tn->azimuth}=\frac{\Vec{m}}{P_n(\phi_{Bob})}
 \label{eq1a}\\
 s_{tn->elevation}=\frac{\Vec{m}}{P_n(\theta_{Bob})}
 \label{eq1b}
\end{align}
\end{subequations}

\noindent where $s_{tn}$ denotes the $n$th transmitted signal, $n=1,2,..5$ represents the port number in the proposed electrically small antenna; $P_n{(\phi_{Bob})}$ and $P_n{(\theta_{Bob})}$ are the complex patterns of the $n$th antenna port in the direction of the legitimate receiver in the azimuth and elevation planes, respectively; $\Vec{m}$ is a complex modulation vector (including artificially generated noise used to scramble the constellation in undesired directions).

To study the DM performance, without loss of generality, Quadrature Phase Shift Keying (QPSK) is used as the modulation scheme. Bit error rate (BER) calculations are executed using $10^5$ transmitted symbols and $\SI{12}{dB}$ Signal to Noise Ratio (SNR) at the intended receiver, with additive white Gaussian noise assumed to be independent for each different location. 

\figref{fig:DM_over_xy} shows the DM performance for the azimuth plane computed using \eqref{eq1a} for different $\phi$ directions covering the entire azimuth plane. It can be seen that the method realizes  steerable transmissions with low BER towards the desired angle $\phi_{Bob}$, without leakage into other eavesdroppers' directions in the same plane. Low BER values around $10^{-5}$ are observed at the desired directions of the legitimate user, and the beamwidth with BER$<10^{-2}$ is around $\ang{55}$.  

\figref{fig:DMPerfm2}a and \figref{fig:DMPerfm2}b show the DM performance for $\phi_{Bob}=\ang{50}$ and $\phi_{Bob}=\ang{180}$ directions. It can be seen, that the communication cannot be intercepted by any eavesdropper located at a different direction $\phi$. The directional modulation performance for the elevation planes is depicted in \figref{fig:DMPerfm}a and \figref{fig:DMPerfm}b for $\theta=\ang{50}$ and $\theta=\ang{120}$, respectively. It is also observed that steerable transmission is realized in the elevation plane with a low BER of $10^{-5}$ at the desired angle $\theta$, and the beamwidth with BER$<10^{-2}$ is around $\ang{58}$. 
Analogously to the previous case, the communication cannot be intercepted by any eavesdropper located at a different direction $\theta$ other than the $\theta_{Bob}$.
This security is realized with highly miniaturized antennas, making the proposed design a promising candidate for size-constrained IoT applications. 

\begin{figure}[ht]
 \vspace{.25cm}
	\subfloat[]{
	\setlength\figureheight{0.19\textwidth}
	\setlength\figurewidth{0.3\textwidth}
	\input{New_2D_Azi50deg}}
    \hfill
	\subfloat[]{
	\setlength\figureheight{0.19\textwidth}
	\setlength\figurewidth{0.3\textwidth}
	\input{New_2D_Azi180deg}}
	\caption{BER calculations using $\SI{12}{dB}$ SNR showing the DM performance in the azimuth plane for (a) $\phi=\ang{50}$; and (b) at $\phi=\ang{180}$.}
	\label{fig:DMPerfm2}

	\subfloat[]{
	\setlength\figureheight{0.19\textwidth}
	\setlength\figurewidth{0.3\textwidth}
	\input{New_2D_Elev50deg}}
    \hfill
	\subfloat[]{
	\setlength\figureheight{0.19\textwidth}
	\setlength\figurewidth{0.3\textwidth}
	\input{New_2D_Elev120deg}}
	\caption{BER calculations using $\SI{12}{dB}$ SNR showing the DM performance in the elevation planes for (a) the receiver located at $\theta=\ang{50}$; and (b) $\theta=\ang{120}$.}
	\label{fig:DMPerfm}
\end{figure}

\section*{Acknowledgment}
This publication has emanated from research conducted with the financial support of Science Foundation Ireland under Grant number $18$/SIRG/$5612$ and in part by the IEEE Antennas and Propagation Society Doctoral Research Grant 2022.

\balance


\begin{thebibliography}{1}

\bibitem{Qadri2020}
Y.~A.~Qadri, A.~Nauman, Y.~B.~Zikria, A.~V.~Vasilakos and S.~W.~Kim, ``The Future of Healthcare Internet of Things: A Survey of Emerging Technologies'', \emph{IEEE Communications Surveys \& Tutorials}, vol. 22, no. 2, pp. 1121-1167, Secondquarter 2020.

\bibitem{Dian2020}
F.~J.~Dian, R.~Vahidnia and A.~Rahmati, ``Wearables and the Internet of Things (IoT), Applications, Opportunities, and Challenges: A Survey,'' \emph{IEEE Access}, vol. 8, pp. 69200-69211, 2020.

\bibitem{Munoz2009}
D.~Munoz,~F.~Bouchereau,~C.~Vargas,~and~R.~Enriquez-Caldera,~\emph{``Position
Location Techniques and Applications.''}~Academic~Press,~2009.

\bibitem{Obeidat2021}
H.~Obeidat,~W.~Shuaieb,~O.~Obeidat,~and~R.~Abd-Alhameed,~``A review of
indoor localization techniques and wireless technologies,''\emph{~Wireless~Pers.
Commun.},~vol.~117,~no.~4,~pp. 1–39, Feb. 2021.

\bibitem{IoTMAG22}
A. Zandamela, A. Chiumento, N. Marchetti and A. Narbudowicz, ``Angle of Arrival Estimation via Small IoT Devices: Miniaturized Arrays vs. MIMO Antennas'', \emph{IEEE Internet of Things Magazine}, vol. 5, no. 2, pp. 146-152, June 2022.

\bibitem{Daly2009}
M.~P.~Daly and J.~T.~Bernhard, ``Directional Modulation Technique
for Phased Arrays,'' \emph{IEEE Trans. Antennas Propag.}, vol. 57, no. 9,pp. 2633–2640, Sep. 2009.

\bibitem{Ding2014}
Y.~Ding and V.~F.~Fusco, ``A Vector Approach for the Analysis and Synthesis of Directional Modulation Transmitters,'' \emph{IEEE Trans. Antennas Propag.}, vol. 62, no. 1, pp. 361–370, Jan. 2014.

\bibitem{Trung2018}
D.~Trung,~Z.~Xiangyun, and~H.~V.~Poor, \emph{``Trusted Communications with
Physical Layer Security for 5G and Beyond'}', Institution of Engineering and Technology, 2018.

\bibitem{Wei2020}
Z.~Wei, C.~Masouros, F.~Liu, S.~Chatzinotas and B.~Ottersten, ``Energy- and Cost-Efficient Physical Layer Security in the Era of IoT: The Role of Interference,'' \emph{IEEE Communications Magazine}, vol. 58, no. 4, pp. 81-87, April 2020.

\bibitem{AWPL2022}
A.~Narbudowicz, A.~Zandamela, N.~Marchetti, and M.~J.~Ammann,
``Energy-Efficient Dynamic Directional Modulation With Electrically Small Antennas,'' \emph{IEEE Antennas and Wireless Propagation Letters,} vol. 21, no. 4, pp. 681–684, 2022.

\bibitem{Iqbal2021}
Z.~Iqbal,~T.~H.~Mitha~and~M.~Pour, ``A Reconfigurable Radiation Pattern Microstrip Patch Antenna with High Mode Purity,''~\emph{ 2021 IEEE International Symposium on Antennas and Propagation and USNC-URSI Radio Science Meeting (APS/URSI)}, 2021, pp. 71-72.

\bibitem{Zhao2022}
S.~Zhao,~Z.~Wang~and~Y.~Dong,~``A Planar Pattern-Reconfigurable Antenna With Stable Radiation Performance,'' \emph{IEEE Antennas Wirel. Propag. Letters}, vol. 21, no. 4, pp. 784-788, April 2022.

\bibitem{Huang2021}
H.~-F.~Huang~and~H.~-l.~Bu,~``Single-Layer Electrically Small Antenna With Reconfigurable Radiation Pattern,''~\emph{2021 IEEE MTT-S International Wireless Symposium (IWS)}, 2021, pp. 1-3.

\bibitem{Wu2020}
Z.~Wu,~M.~-C.~Tang,~M.~Li~and~R.~W.~Ziolkowski,~``Ultralow-Profile, Electrically Small, Pattern-Reconfigurable Metamaterial-Inspired Huygens Dipole Antenna,''~\emph{IEEE Trans. Antennas Propag.}, vol. 68, no. 3, pp. 1238-1248, March 2020.

\bibitem{AbelAEU}
A. Zandamela, K. Schraml, S. Chalermwisutkul, D. Heberling, and A. Narbudowicz,~``Digital pattern synthesis with a compact MIMO antenna of half-wavelength diameter,''~\emph{AEU – Int. J. Electron. Commun.},~vol.~135,~Jun.~2021.

\bibitem{Abel2019}
A. A. Zandamela et al., ``On the Efficiency of Miniaturized 360° Beam-Scanning Antenna,'' \emph{2019 13th European Conference on Antennas and Propagation (EuCAP)}, 2019, pp. 1-4.
\end{thebibliography}
\end{document}